\documentclass[11pt,a4paper]{article}

\usepackage[utf8]{inputenc}
\usepackage[margin=1in]{geometry}
\usepackage{amsmath,amssymb}
\usepackage[authoryear]{natbib}
\usepackage{graphicx}
\usepackage{booktabs}
\usepackage{algorithm}
\usepackage{algpseudocode}
\usepackage{color}
\usepackage{setspace}
\usepackage{soul}
\usepackage{hyperref}

\title{GreenPhase: A Green Learning Approach for Earthquake Phase Picking}

\author{
Yixing Wu$^{1}$,
Shiou-Ya Wang$^{2}$,
Dingyi Nie$^{1}$,
Sanket Kumbhar$^{1}$,
Yun-Tung Hsieh$^{3}$,\\
Yun-Cheng Wang$^{1}$,
Po-Chyi Su$^{3}$,
C.-C. Jay Kuo$^{1}$\\[0.5em]
\small $^{1}$Ming Hsieh Department of Electrical and Computer Engineering, University of Southern California\\
\small $^{2}$Center for Earth and Environmental Studies, National Central University\\
\small $^{3}$Department of Computer Science \& Information Engineering, National Central University\\[0.25em]
\small \texttt{yixingwu@usc.edu}
}

\date{}

\begin{document}

\maketitle

\begin{abstract}
Earthquake detection and seismic phase picking are fundamental yet challenging tasks in seismology due to low signal-to-noise ratios, waveform variability, and overlapping events. Recent deep-learning models achieve strong results but rely on large datasets and heavy backpropagation training, raising concerns over efficiency, interpretability, and sustainability.

We propose GreenPhase, a multi-resolution, feed-forward, and mathematically interpretable model based on the Green Learning framework. GreenPhase comprises three resolution levels, each integrating unsupervised representation learning, supervised feature learning, and decision learning. Its feed-forward design eliminates backpropagation, enabling independent module optimization with stable training and clear interpretability. Predictions are refined from coarse to fine resolutions while computation is restricted to candidate regions.

On the Stanford Earthquake Dataset (STEAD), GreenPhase achieves excellent performance with F1 scores of 1.0 for detection, 0.98 for P-wave picking, and 0.96 for S-wave picking. This is accomplished while reducing the computational cost (FLOPs) for inference by approximately 83\% compared to state-of-the-art models. These results demonstrate that the proposed model provides an efficient, interpretable, and sustainable alternative for large-scale seismic monitoring.
\end{abstract}

\noindent\textbf{Keywords:} Earthquake detection; Phase picking; Multi-resolution analysis; Green Learning; Interpretable machine learning; Sustainable computing

\section{Introduction}
\label{intro}

Earthquake detection and phase picking are core tasks in seismology, underpinning the analysis of seismic waves, accurate hypocenter location, imaging of Earth’s interior, and applications such as earthquake early warning. Accurate determination of these parameters is crucial for subsequent analyses, yet it remains challenging in practice.

Initially, this task relied on manual interpretation by experienced analysts. However, this method takes a lot of time and manual effort, and it can lead to mistakes because of human judgment. Early studies introduced automated tools such as STA/LTA~\cite{allen1978automatic}, the Akaike Information Criterion (AIC)~\cite{takanami1988new}, and cross-correlation~\cite{vandecar1990determination, gibbons2006detection}. While these methods accelerated processing, they still had significant drawbacks. They were highly dependent on signal quality, often leading to misjudgments in low signal-to-noise ratio environments, and their parameter settings were not universally applicable, making them difficult to adapt across different regions.~\cite{withers1998comparison}

To improve efficiency, Deep Learning (DL)~\cite{lecun2015deep} has been introduced into seismology in recent years. Several DL-based methods, including EQTransformer~\cite{mousavi2020earthquake}, PhaseNet~\cite{zhu2019phasenet}, GPD~\cite{ross2018generalized}, PickNet~\cite{wang2019deep}, PpkNet~\cite{zhou2019hybrid}, and Yews~\cite{zhu2019deep}, have shown strong performance in earthquake detection and seismic phase picking. These models learn the general characteristics of earthquake waveforms and seismic phases from high-level representations. Therefore, compared with traditional methods, they can adapt to variable geological settings and process data much faster. Despite their success, DL approaches rely heavily on backpropagation, which raises sustainability concerns due to high computational demands and associated carbon footprints. Moreover, their black-box nature makes it difficult to interpret the underlying mechanisms in a mathematically rigorous way.

To overcome these limitations, we propose GreenPhase, which leverages the Green Learning (GL) framework~\cite{kuo2023green}—an emerging paradigm that improves efficiency, explainability, and generalizability. Compared with conventional deep neural networks, GL offers three key advantages: (1) reduced computational complexity, (2) robust performance with fewer training samples, and (3) improved mathematical and statistical transparency. This makes GL particularly attractive for applications such as seismology, where scalability and interpretability are critical.

The Green Learning framework has been successfully applied to various machine learning tasks, including image classification~\cite{chen2020pixelhop, chen2020pixelhop++}, segmentation~\cite{yang2025gusl}, restoration~\cite{11084379} and medical analysis~\cite{wu2025prediction}. Owing to the structural similarity between time sequences and images, this framework can also be extended to seismic phase picking, where it has demonstrated both efficiency and effectiveness.

Unlike traditional deep neural networks that rely on computational neurons and end-to-end backpropagation, GL adopts a structured feed-forward design. The framework consists of three sequential modules: (1) unsupervised representation learning, (2) supervised feature learning, and (3) supervised decision learning. Each stage produces intermediate outputs that become the inputs to the next, so parameters are learned locally within a module rather than via gradients flowing through the entire pipeline. The final decision module is trained independently, without backpropagating through earlier stages. This feed-forward structure brings practical advantages, such as lower computational and energy cost (no multi-epoch end-to-end tuning), and fewer and less sensitive hyperparameters. It also enhances interpretability, as each module serves a clear purpose. In seismology, this translates into efficient large-scale processing and transparent reasoning behind each phase pick.

To further enhance efficiency, we adopt a hierarchical multi-resolution scheme in which predictions proceed from the coarsest to the finest temporal resolution. The coarsest level operates on a downsampled version of the seismic sequence to propose candidate time locations with a high likelihood of containing P- or S-wave onsets. Higher-resolution levels then re-examine only these locations, applying the same three GL modules—unsupervised representation learning, supervised feature learning, and supervised decision learning—to progressively refine the predictions. This coarse-to-fine schedule avoids scanning the entire record at full resolution and concentrates computation on the most informative portions of the signal.

In this study, we trained GreenPhase, a GL-based model for earthquake detection and seismic phase picking, using the widely adopted Stanford Earthquake Dataset (STEAD)~\cite{mousavi2019stanford}. We benchmark our method against state-of-the-art deep learning and traditional approaches. The results demonstrate that GreenPhase achieves comparable performance while requiring significantly fewer computational resources, underscoring its potential as a sustainable and interpretable alternative for large-scale seismological analysis.

\section{Methodology}

\subsection{Preprocessing}
We used the Stanford Earthquake Dataset (STEAD)~\cite{mousavi2019stanford} to train and evaluate GreenPhase. STEAD is a large-scale, globally distributed database of labeled earthquake and non-earthquake waveforms. STEAD includes over one million earthquake recordings together with noise records collected at seismic stations with epicentral distances up to 300 km.

Each waveform in STEAD is provided as a one-minute, three-component seismic sequence sampled at 100 Hz and band-pass filtered between 1.0 and 45.0 Hz. At the sampling rate of 100 Hz, each seismic waveform is represented as 
\[
\mathbf{X} \in \mathbb{R}^{3 \times L}, \quad L = 6000,
\]
where the three rows correspond to the E, N, and Z components, and $L$ denotes the number of time samples per channel. Earthquake records are annotated with P- and S-wave arrival times, whereas noise records contain no phase picks. For fair comparison with existing deep-learning methods such as EQTransformer~\cite{mousavi2020earthquake}, we adopted the same testing partition protocol utilized in this work, ensuring that our evaluation is directly comparable to previously reported results.

\begin{figure}
  \centering
  \includegraphics[width=0.85\textwidth]{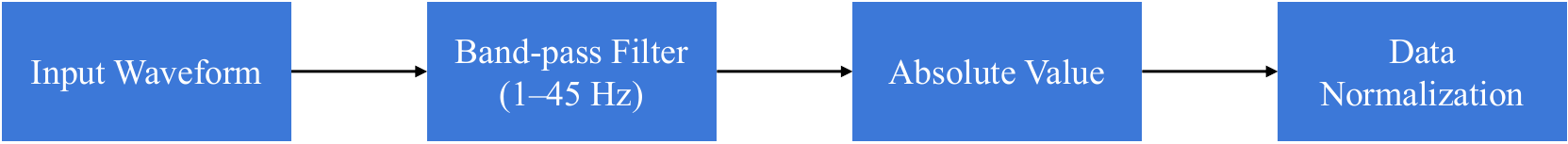}
  \caption{Data preprocessing workflow prior to GreenPhase training.}
  \label{fig:preprocess_pipeline}
\end{figure}

Figure~\ref{fig:preprocess_pipeline} illustrates the general data preprocessing workflow. The first step involves applying a band-pass filter (1-45 Hz) to the input data. The band-pass filter is employed to eliminate very low-frequency noise while preserving the valuable signal content from seismic waves. Since P and S waves primarily contain frequencies within the 1-45 Hz range, the band-pass filter effectively removes irrelevant frequencies. Frequencies above 45 Hz often carry noise from sources like nearby machinery, urban activity, or surface waves, which can obscure important seismic signals. By removing these high-frequency components, the band-pass filter enhances the clarity of the recorded waveforms.

After band-pass filtering, we take the absolute value of each waveform to emphasize signal magnitude. Then, for every waveform, the maximum and minimum values are determined across all three channels (E, N, and Z) and the entire time sequence ($L=6000$ samples). 
Each waveform is then normalized to the range $[0,1]$ using
\[
\hat{\mathbf{X}} = \frac{|\mathbf{X}| - \min(|\mathbf{X}|)}{\max(|\mathbf{X}|) - \min(|\mathbf{X}|)}.
\]

This normalization ensures that the maximum amplitude in each sample is scaled to 1, while the minimum is set to 0. Waveforms for which the maximum and minimum are equal (e.g., constant signals or traces dominated by zeros after filtering) are discarded, since they contain no useful information. 

\begin{figure}
  \centering
  \includegraphics[width=0.85\textwidth]{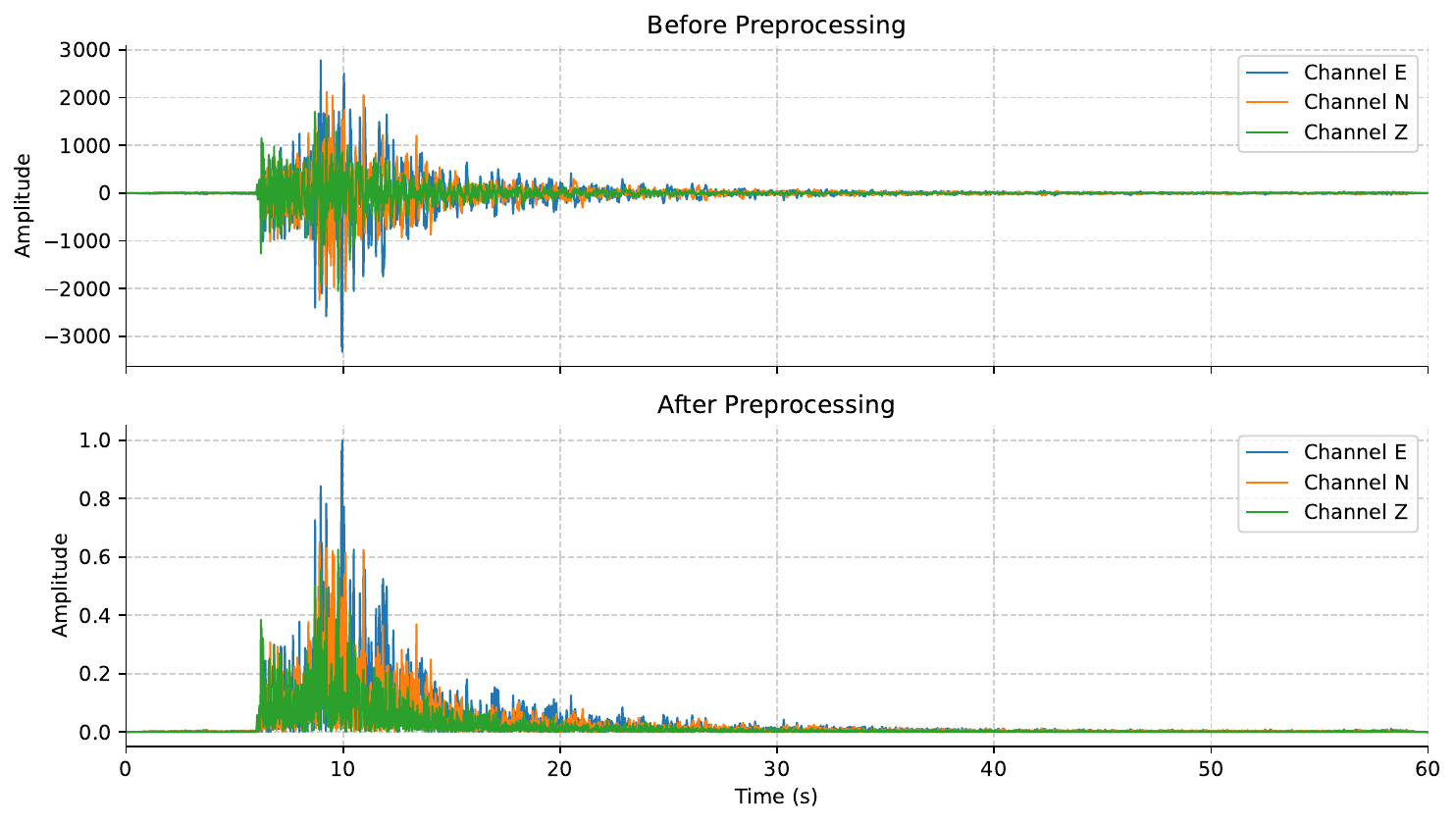}
  \caption{Comparison of a seismic waveform before and after the preprocessing pipeline.}
  \label{fig:pre_waveform}
\end{figure}

After this step, the normalized waveform
$\hat{\mathbf{X}} \in \mathbb{R}^{3 \times L}$
is used as the input to the GreenPhase. By constraining all records to the same dynamic range, this preprocessing removes amplitude bias across different events, stabilizes feature extraction, and improves the robustness of the subsequent learning modules. Figure~\ref{fig:pre_waveform} illustrates the effect of normalization by comparing input data before and after processing. 

\subsection{Multi-resolution Design}
The GreenPhase consists of three main components: P-wave picking, S-wave picking, and seismic event detection. Each input waveform is first processed by the P-wave and S-wave picking modules to predict the respective arrival times. Based on these predictions and their associated probabilities, the seismic detection module then determines whether the input waveform corresponds to an actual earthquake or to noise.  

\begin{figure*}[t]
  \centering
  \includegraphics[width=\textwidth]{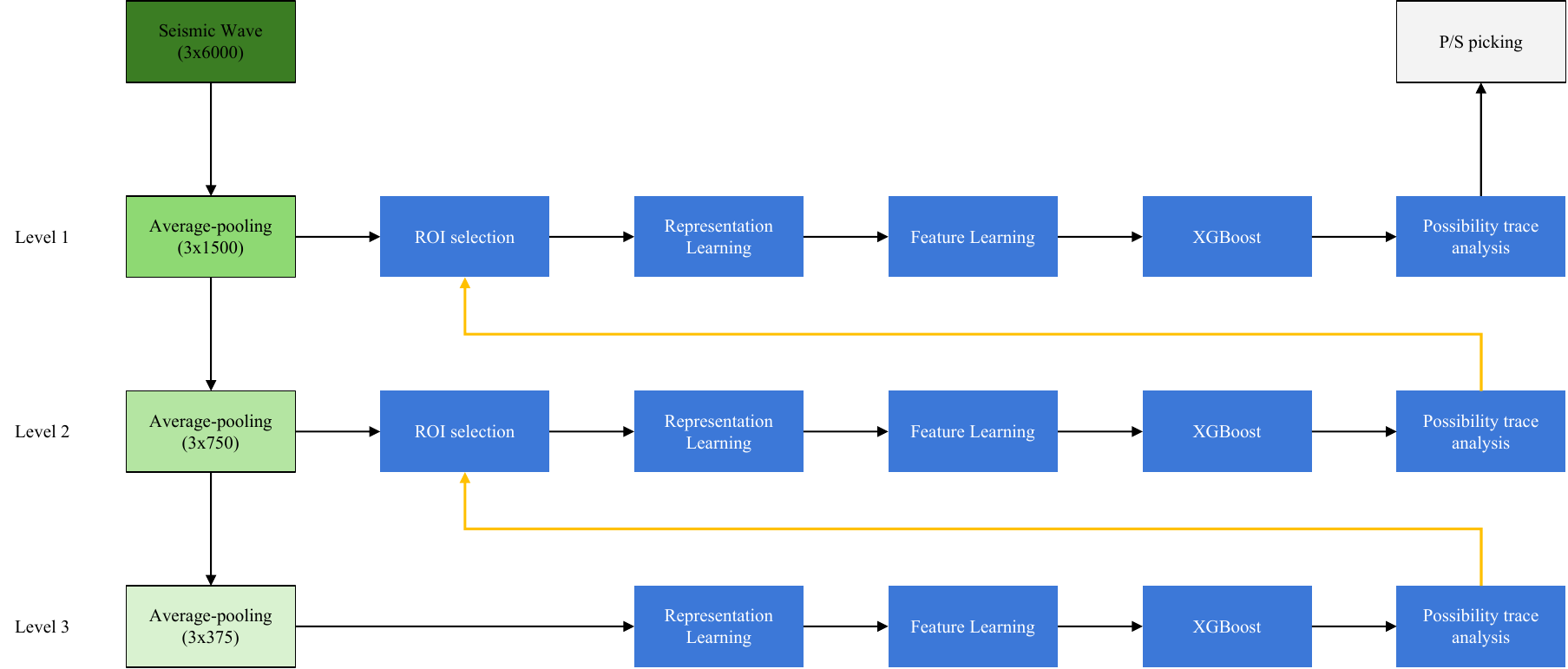}
  \caption{The architecture of P/S-wave picking}
  \label{fig:P/S-wave pipeline}
\end{figure*}

We first describe the architecture of the P- and S-wave picking modules, which share the same design and are illustrated in Figure~\ref{fig:P/S-wave pipeline}. Each waveform in the STEAD dataset has a dimension of $3 \times 6000$, which can be treated as a two-dimensional input array. To capture local temporal patterns, we adopt a multi-resolution framework. From level 1 to level 3, the temporal resolution is progressively reduced through average pooling from 6000 to 1500, 750, and 375 samples. At the coarsest level, GreenPhase evaluates every time location to compute the probability of P- and S-wave arrivals. In higher levels, however, the model no longer scans the entire sequence; instead, it focuses only on the candidate regions indicated by the previous coarser level. This hierarchical design allows GreenPhase to first analyze coarse-scale information and then refine its predictions at finer resolutions, substantially reducing the computational resources required during inference. Finally, the arrival time predicted at the finest resolution (1500 samples) is scaled back to the original 6000-sample coordinate system to produce the final pick.

At each resolution level, the input waveform is first processed by the Saab (Subspace Approximation via Adjusted Bias) transform~\cite{kuo2019interpretable} to extract spatial–spectral representations. For every time location within the candidate regions identified at the previous resolution level, the Saab transform extracts a set of features that capture both temporal and spectral patterns. 

These features are then passed to the supervised learning modules. Specifically, the Relevant Feature Test (RFT)~\cite{yang2022supervised} is employed to select the most discriminative features, while the Statistics-based Feature Generation (SFG)~\cite{wang2024statistics} produces additional features from the raw representations. Finally, an XGBoost regressor~\cite{chen2016xgboost} is used to predict the probability that the center time location corresponds to a P- or S-wave arrival.

\subsection{Label Generation}

The objective of training is to assign a continuous-valued pseudo-label, denoted as $y_i$, to each time index $i$ of the input waveform. This label represents the likelihood of a P- or S-wave arrival at that location.

To incorporate local temporal context, we define a symmetric window $w_i$ centered at each time index $i$. The window is defined by the interval $[i-h_k, i+h_k]$, where $h_k$ is the window half-width at resolution level $k$. For levels $k \in \{1,2,3\}$, the half-widths $h_k$ are 64, 32, and 16 samples, respectively. This window definition is consistently used for both (i) applying the Saab transform to the window $w_i$ to extract features, and (ii) assigning the supervision pseudo-label $y_i$ to the center index $i$. This alignment ensures that the training signal matches the model’s effective receptive field at every resolution.

Let $p$ be the ground-truth arrival time. The value of the pseudo-label $y_i$ is calculated based on the position of $p$ relative to the window $w_i$. A label of $y_i = 1.0$ is assigned if the arrival is within a small tolerance of the center, i.e., $|p-i| \leq \Delta t_k$. For levels 1, 2, and 3, the corresponding tolerances $\Delta t_k$ are 12, 6, and 3 samples, respectively. If the arrival is inside the window ($p \in [i-h_k, i+h_k]$) but outside this tolerance, the label is computed using the following regression formula:

\begin{equation}
\label{eq:label}
y_i = \frac{\min(L,R)}{\max(L,R)}
\end{equation}

Here, $L = p-(i-h_k)$ is the distance from the arrival to the window's start, and $R = (i+h_k)-p$ is the distance to the window's end. For noise events where no arrival exists, $y_i$ is set to zero for all $i$.

\begin{figure}[t]
    \centering
    \includegraphics[width=\columnwidth]{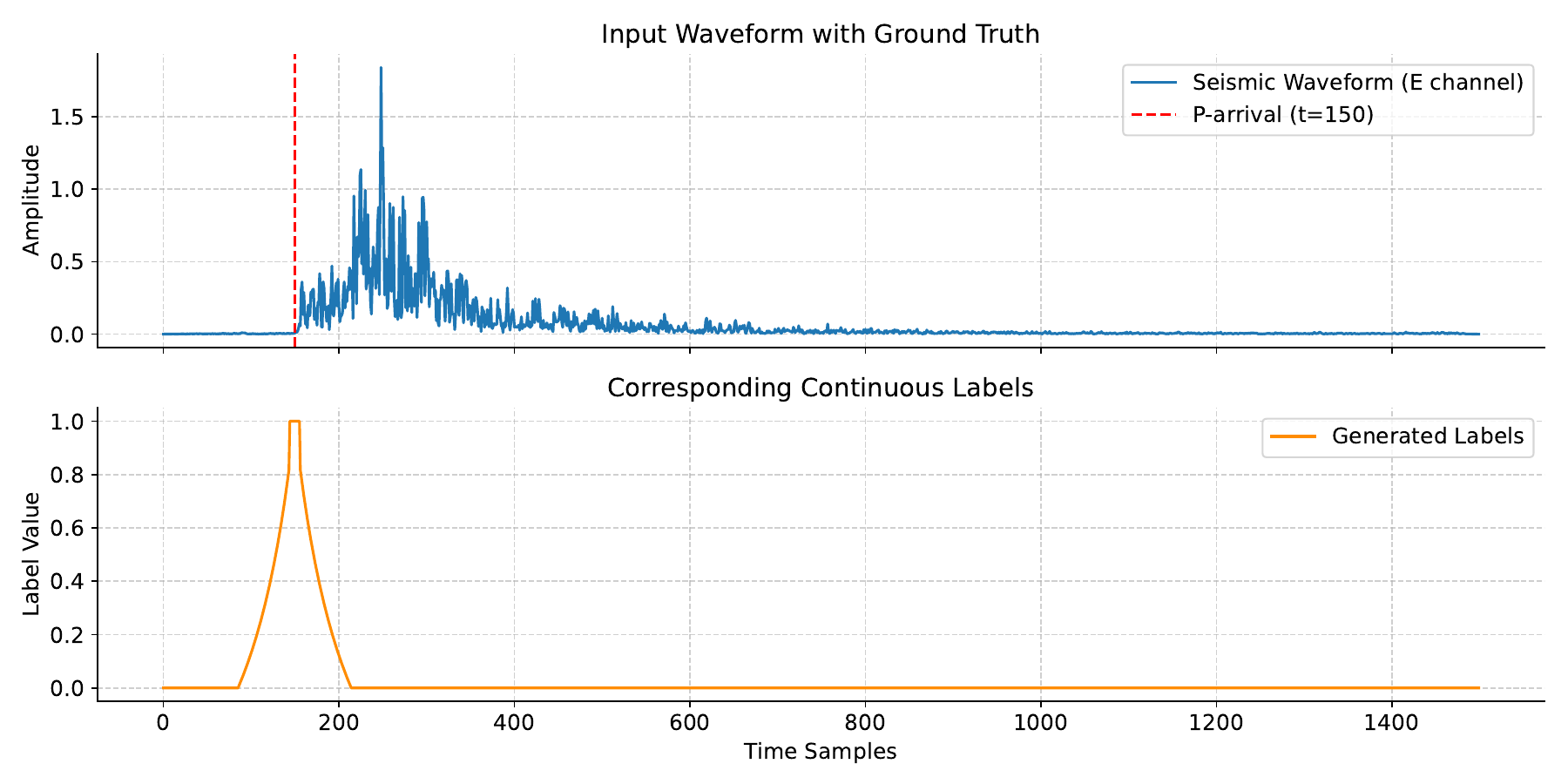}
    \caption{An example of the pseudo-label generation for a Level 1 waveform (1500-sample resolution). The top panel shows the transformed waveform with the P-arrival marked. The bottom panel displays the corresponding continuous pseudo-label for each time location.}
    \label{fig:label_across_seq}
\end{figure}

Figure~\ref{fig:label_across_seq} illustrates the pseudo-label generation process for a waveform at Level 1 (1500 samples). The top panel displays the pre-processed waveform, with the ground-truth P-arrival location marked by the red dashed line. The bottom panel shows the corresponding continuous-valued pseudo-labels. This label peaks at the arrival time and smoothly decays, providing a focused and continuous supervision signal for the model.

\subsection{Sampling}
After generating the pseudo-label $y_i$ for each time index $i$, the resulting dataset of all windows $\{w_i\}$ is heavily skewed toward those with zero labels ($y_i=0$). To address this imbalance, we apply a stratified subsampling strategy. Each window $w_i$ is grouped into one of three categories based on its label $y_i$: high-confidence ($y_i \geq 0.8$), intermediate ($0 < y_i < 0.8$), and noise ($y_i=0$). The final training set is then constructed by randomly drawing an equal number of windows from each of the three categories, ensuring a balanced dataset and improving the model's robustness.

\begin{figure}[h]
\centering
\includegraphics[width=\columnwidth]{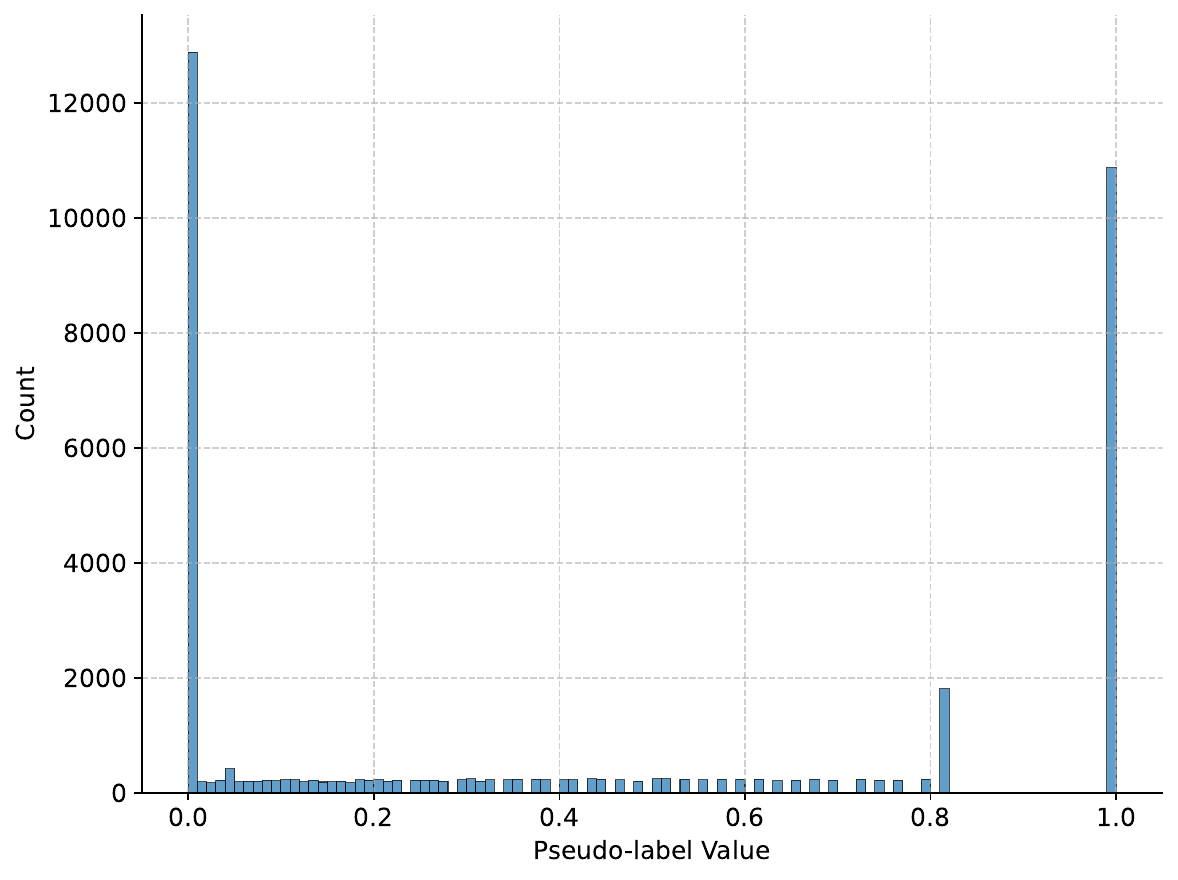}
\caption{The distribution of pseudo-label values for the Level 1 training set \textbf{after} stratified sampling. The histogram confirms that the three categories (noise, intermediate, and high-confidence) are equally represented, resulting in a balanced dataset for model training.}
\label{fig:level_1_pseudo_labels}
\end{figure}

The result of this balancing process for the Level 1 training data is illustrated in Figure~\ref{fig:level_1_pseudo_labels}. The histogram shows the distribution of pseudo-label values after stratified sampling has been applied. As shown, the three categories are now equally represented, creating a balanced distribution that prevents the model from being biased toward the initially dominant noise class.

\subsection{Representation Learning}

The Saab transform~\cite{kuo2019interpretable} is a variation of Principal Component Analysis (PCA)~\cite{wold1987principal} developed within the GL framework. It is designed to capture features from both near and distant neighborhoods of the input waveform without label assistance, thereby extracting joint spatial–spectral representations at the center time location of each window. Unlike standard PCA, the Saab transform first computes the Direct Current (DC) component, i.e., the mean of the data, and then applies PCA to the DC-reduced data to extract Alternating Current (AC) components, resulting in a richer set of features.

In our multi-resolution framework, the Saab transform is applied to each window $w_i$ at each level $k \in \{1,2,3\}$. We use sliding kernels of varying sizes to capture both fine- and coarse-scale temporal patterns within the window. For levels 1 and 2, we use a kernel of size $3 \times 16$ with a stride of 8; for level 3, the kernel size is $3 \times 8$ with a stride of 4. This multi-scale kernel approach allows the Saab transform to function as a data-driven filter bank, where larger kernels capture longer-range dependencies and smaller kernels capture local variations. The resulting features form a comprehensive set of candidates for the subsequent feature learning stage.

In addition to the Saab features, we design a window energy feature to capture local energy asymmetry around the arrival time. For a window centered at index $i$ with half-width $h_k$, the window is divided into two parts: the left half $[i-h_k,\, i-1]$ and the right half $[i+1,\, i+h_k]$. We compute the average squared amplitude (signal power) in each half and then take their difference, further averaged across all channels. This value, referred to as the window energy, reflects whether the signal power is stronger before or after the center, effectively highlighting abrupt energy changes that indicate potential P- or S-wave arrivals.

\subsection{Feature Learning}
The large number of features produced during representation learning cannot all be directly passed to the decision module, as this would cause unnecessary computational overhead. To address this, we employ the RFT~\cite{yang2022supervised} to identify the most informative and discriminative features under supervision, thereby reducing the feature space.

RFT evaluates the relevance of each feature dimension to the supervised target by estimating its RFT loss. Specifically, the value range of a feature is uniformly divided into bins, and the midpoint of each bin is considered as a candidate threshold. Each threshold is then used to split the data into two subsets, and the weighted mean squared error (MSE) of the target is computed for the partition. The lowest MSE across all thresholds is taken as the RFT loss of that feature, reflecting its discriminative capacity. Features with smaller RFT loss values are considered more informative, and the effective set of features is determined by selecting those before the elbow point of the ascending RFT loss curve. 

Beyond feature selection, we further enrich the representation space using the Statistics-based Feature Generation (SFG) method~\cite{wang2024statistics}. This method leverages the supervision of residuals to identify subsets of discriminative features and combines them into new features via linear projection. The process consists of two steps: (1) selecting representation subsets with the help of a shallow XGBoost regressor~\cite{chen2016xgboost}, which partitions the target space and groups features along each tree path; and (2) generating new features through the Least-Square Normal Transform (LNT)~\cite{10386982}, which linearly projects each subset to the target subspace. The resulting LNT features are more directly related to the regression target, providing stronger discriminative power for the final decision stage.

\subsection{Decision Learning for P/S-wave Picking}

The final decision-making stage is a hierarchical, coarse-to-fine procedure that begins at the coarsest level (Level 3). At this initial level, the XGBoost regressor is applied to the entire sequence to produce a probability trace over time. From this trace, we locate the precise arrival time using a neighborhood analysis: we find the global maximum value of the trace, define a threshold at 95\% of this maximum, and select the first peak that exceeds this threshold as the predicted arrival.

This prediction is then scaled up to the coordinate system of the next finer level, where it defines a narrow Region of Interest (ROI) spanning $\pm 40$ samples around the scaled prediction. At the finer levels, the model's analysis is constrained to this small ROI, drastically reducing computation. The same peak-finding procedure is repeated on the new, localized probability trace to refine the arrival time. This refinement process is repeated up to the finest resolution level.

For S-phase picking at each level, the process is identical, but the probability trace is first masked by setting all values prior to that level's predicted P-phase arrival to zero. This entire hierarchical approach ensures both computational efficiency and high-resolution accuracy.

\subsection{Seismic Wave Detection}
After the P/S-wave picking stage, the model outputs the predicted arrival times of the P- and S-waves along with their associated probabilities across the three resolution levels. An additional task is to determine whether the input waveform corresponds to a real seismic event or to noise. For this purpose, we train a lightweight XGBoost classifier. The input to this classifier includes, for both the P-wave and S-wave models at each of the three resolution levels, the predicted arrival time location and its associated probability. This yields four features per level, resulting in 12 features in total. In addition, we include the distance between the final predicted P- and S-wave arrival times, giving a 13-dimensional feature vector that is used to train the XGBoost classifier. The XGBoost classifier is lightweight, with a maximum depth of 2 and using a decision threshold of 0.5.

\section{Results}

\subsection{Dataset}
The GreenPhase model is trained and evaluated using the Stanford Earthquake Dataset~\cite{mousavi2019stanford}. For fair comparison with existing deep-learning and traditional approaches, we adopt the same testing set as EQTransformer~\cite{mousavi2020earthquake}, which includes over 120,000 waveforms consisting of both earthquake and noise examples for evaluating detection and picking performance. The remaining portion of the dataset, containing approximately 1.2 million seismic events, is used for training and validation. We further split this set into 80\% for training and 20\% for validation.

\subsection{Detection Performance}
As summarized in Table~\ref{tab:detection_performance}, our method achieves state-of-the-art results on the seismic event detection task. Using a simple XGBoost classifier with a decision threshold of 0.5 and features derived from our GL-based P- and S-wave picking models, it matches or surpasses the performance of existing approaches such as EQTransformer, CRED, DetNet, Yews, and STA/LTA, despite relying on relatively lightweight input features.

\begin{table*}[t]
\centering
\caption{Detection performance.}
\label{tab:detection_performance}
\begin{tabular}{lcccccc}
\toprule
\textbf{Model} & \textbf{Pr} & \textbf{Re} & \textbf{F1} & \textbf{Training size} & \textbf{Ref.} \\
\hline
GreenPhase (Ours) & \textbf{1.0} & \textbf{1.0} & \textbf{1.0}  & 1.2M & This study \\
EQTransformer & \textbf{1.0} & \textbf{1.0} & \textbf{1.0}  & 1.2M & ~\cite{mousavi2020earthquake} \\
CRED          & \textbf{1.0} & 0.96         & 0.98         & 1.2M & ~\cite{mousavi2019cred} \\
DetNet        & \textbf{1.0} & 0.89         & 0.94         & 30K  & ~\cite{zhou2019hybrid} \\
Yews          & 0.84         & 0.85         & 0.85         & 1.4M & 
~\cite{zhu2019deep} \\
STA/LTA       & 0.91         & \textbf{1.0} & 0.95         & ---  & ~\cite{allen1978automatic} \\
\bottomrule
\end{tabular}
\end{table*}

\subsection{P/S Wave Picking Performance}
We evaluate the performance of P- and S-phase picking using precision (Pr), recall (Re), and F1, and compare against both deep-learning and traditional baselines. The calculation of these metrics explicitly incorporates both detection correctness and timing accuracy, with a fixed tolerance of 0.5 s. Because each waveform in STEAD is a one-minute, three-component seismic record sampled at 100 Hz, this corresponds to 50 time samples.

A prediction is counted as a true positive (TP) if it correctly identifies a ground-truth phase arrival and its absolute timing error does not exceed 0.5 s. A false positive (FP) arises in two situations: (1) when the model predicts a phase arrival in a noise waveform where no ground-truth arrival exists, or (2) when the prediction corresponds to a true arrival but its timing error is greater than 0.5 s, i.e., the pick is imprecise. A false negative (FN) occurs when a ground-truth phase arrival is present in the waveform, but the model fails to produce a corresponding prediction within 0.5 s. Under these definitions, precision, recall, and F1 provide a balanced evaluation of both the ability to correctly detect phases and the accuracy of their predicted arrival times.

Based on these definitions, the evaluation metrics are computed as:

\begin{equation}
\text{Precision} = \frac{\text{TP}}{\text{TP} + \text{FP}}
\end{equation}
\begin{equation}
\text{Recall} = \frac{\text{TP}}{\text{TP} + \text{FN}}
\end{equation}
\begin{equation}
\text{F1} = 2 \times \frac{\text{Precision} \times \text{Recall}}{\text{Precision} + \text{Recall}}
\end{equation}

The P-phase picking performance of our GL model is evaluated against nine existing methods, including six deep-learning approaches and three traditional techniques, with results summarized in Table~\ref{tab:p_phase_picking}. Our model achieves Pr=0.96, Re=0.99, and F1=0.98. This performance is close to the state-of-the-art benchmark set by EQTransformer (Pr=0.99, Re=0.99, F1=0.99)~\cite{mousavi2020earthquake}. While EQTransformer establishes the state of the art with an F1 of 0.99, our model reaches a competitive F1 of 0.98, ranking as the second-best performer in this comprehensive comparison. Notably, even with far fewer training samples, our method achieves the highest recall and a strong overall F1 score, surpassing multiple deep-learning baselines such as PhaseNet (F1=0.96) and PpkNet (F1=0.90), as well as traditional approaches like Kurtosis, FilterPicker, and AIC. Furthermore, our ablation experiments show that reducing the training set size from 1.2M to 240K or even 60K waveforms still yields F1 scores of 0.98 and 0.97, respectively.

\begin{table*}[t]
\centering
\caption{P-phase picking performance.}
\label{tab:p_phase_picking}
\begin{tabular}{lccccc}
\toprule
\textbf{Model} & \textbf{Pr} & \textbf{Re} & \textbf{F1} & \textbf{Training size} & \textbf{Ref.} \\
\hline
GreenPhase (Ours) & 0.96 & \textbf{0.99} & 0.98 & 1.2M  & This study \\
GreenPhase (Ours) & 0.96 & \textbf{0.99} & 0.98 & 240K  & This study \\
GreenPhase (Ours) & 0.95 & \textbf{0.99} & 0.97 & 60K  & This study \\
EQTransformer & \textbf{0.99} & \textbf{0.99} & \textbf{0.99} & 1.2M  & ~\cite{mousavi2020earthquake} \\
PhaseNet      & 0.96          & 0.96          & 0.96          & 780K  & ~\cite{zhu2019phasenet} \\
GPD           & 0.81          & 0.80          & 0.81          & 4.5M  & ~\cite{ross2018generalized} \\
PickNet       & 0.81          & 0.49          & 0.61          & 740K  & ~\cite{wang2019deep} \\
PpkNet        & 0.90          & 0.90          & 0.90          & 30K   & ~\cite{zhou2019hybrid} \\
Yews          & 0.54          & 0.72          & 0.61          & 1.4M  & ~\cite{zhu2019deep} \\
Kurtosis      & 0.94          & 0.79          & 0.86          & ---   & ~\cite{saragiotis2002pai} \\
FilterPicker  & 0.95          & 0.82          & 0.88          & ---   & ~\cite{lomax2012automatic} \\
AIC           & 0.92          & 0.83          & 0.87          & ---   & ~\cite{maeda1985method} \\
\bottomrule
\end{tabular}
\end{table*}

Similarly, Table~\ref{tab:s_phase_picking} presents the results for S-phase picking. GreenPhase ranks as the second-best overall, following EQTransformer, and consistently achieves the highest recall of 0.99 across all training sizes, even with only 60K training samples. Although EQTransformer attains the best F1 score of 0.98, our model delivers a competitive F1 of 0.96, clearly outperforming several other deep-learning approaches such as PhaseNet (F1=0.94), GPD (F1=0.82), PickNet (F1=0.75), and Yews (F1=0.66), as well as traditional baselines including Kurtosis (F1=0.55), FilterPicker (F1=0.49), and AIC (F1=0.64). For a direct comparison, the baseline performance metrics on detection, P-wave, and S-wave picking for all traditional and deep-learning models were taken from the original EQTransformer study~\cite{mousavi2020earthquake}.

\begin{table*}
\centering
\caption{S-phase picking performance.}
\label{tab:s_phase_picking}
\begin{tabular}{lccccc}
\toprule
\textbf{Model} & \textbf{Pr} & \textbf{Re} & \textbf{F1} & \textbf{Training size} & \textbf{Ref.} \\
\hline
GreenPhase (Ours)          & 0.93          & \textbf{0.99} & 0.96 & 1.2M  & This study \\
GreenPhase (Ours) & 0.93 & \textbf{0.99} & 0.96 & 240K  & This study \\
GreenPhase (Ours) & 0.92 & \textbf{0.99} & 0.95 & 60K  & This study \\
EQTransformer & 0.99          & 0.96          & \textbf{0.98} & 1.2M  & ~\cite{mousavi2020earthquake} \\
PhaseNet      & 0.96          & 0.93          & 0.94          & 780K  & ~\cite{zhu2019phasenet} \\
GPD           & 0.81          & 0.83          & 0.82          & 4.5M  & ~\cite{ross2018generalized} \\
PickNet       & 0.75          & 0.75          & 0.75          & 740K  & ~\cite{wang2019deep} \\
PpkNet        & \textbf{1.00} & 0.91          & 0.95          & 30K   & ~\cite{zhou2019hybrid} \\
Yews          & 0.83          & 0.55          & 0.66          & 1.4M  & ~\cite{zhu2019deep} \\
Kurtosis      & 0.89          & 0.39          & 0.55          & ---   & ~\cite{saragiotis2002pai} \\
FilterPicker  & 0.61          & 0.41          & 0.49          & ---   & ~\cite{lomax2012automatic} \\
AIC           & 0.87          & 0.51          & 0.64          & ---   & ~\cite{maeda1985method} \\
\bottomrule
\end{tabular}
\end{table*}

Table~\ref{tab:flops} summarizes the model size and computational cost of our GreenPhase configurations against EQTransformer. All FLOPs are reported for inference on a single $3 \times 6000$ seismic sequence. When trained on the same 1.2M-waveform dataset, our model has a comparable parameter count to EQTransformer (381K vs. 377K) but is dramatically more efficient, requiring only 22M FLOPs—a nearly 6-fold reduction compared to EQTransformer's 129M FLOPs. The table also demonstrates our model's scalability, with the 60K-sample version offering a highly compact option at just 93K parameters and 4M FLOPs. The EQTransformer model size was computed using the implementation in SeisBench~\cite{woollam2022seisbench}.

\begin{table*}
\centering
\caption{Comparison of model size and testing FLOPs.}
\label{tab:flops}
\begin{tabular}{lccc}
\toprule
\textbf{Model} & \textbf{Training Size} & \textbf{Model Size} & \textbf{FLOPs}\\
\midrule
GreenPhase (Ours) & 60K   & \textbf{93K ($\times$1.0)}   & \textbf{4M ($\times$1.0)}\\
GreenPhase (Ours) & 240K  & 234K (2.5$\times$) & 12M (3.0$\times$)\\
GreenPhase (Ours) & 1.2M  & 381K (4.1$\times$) & 22M (5.5$\times$)\\
\midrule
EQTransformer & 1.2M & 377K (4.1$\times$) & 129M (32.3$\times$)\\
\bottomrule
\end{tabular}
\end{table*}

Beyond its computational efficiency, GreenPhase offers a superior sustainability profile during both training and inference, as detailed in Table~\ref{tab:carbon_test} and Table~\ref{tab:carbon_train}. For the training phase, our largest model was trained on a 16-core AMD EPYC 7513 CPU, completing the task in 8 hours. When compared against the results from the hardware reported in the original EQTransformer paper~\cite{mousavi2020earthquake}, the GreenPhase training process is approximately 40 times more sustainable. For the inference phase, GreenPhase processed over 120,000 test waveforms in 18 minutes, using the 16-core AMD CPU. The EQTransformer benchmark, in contrast, was run on a system equipped with an Intel Core i9-9900K CPU and an NVIDIA 3090 GPU. These results demonstrate that the proposed model provides an efficient, interpretable, and sustainable alternative for large-scale seismic monitoring.

\begin{table*}[t]
\centering
\caption{Comparison of carbon footprint and energy consumption during inference on over 120,000 test waveforms.}
\label{tab:carbon_test}
\begin{tabular}{lccc}
\toprule
\textbf{Model} & \textbf{Training Size} & \textbf{CO\textsubscript{2}e (g)} & \textbf{Energy (Wh)} \\
\midrule
GreenPhase (Ours) & 60K & \textbf{10.30 (1.0$\times$)} & \textbf{24.28 (1.0$\times$)} \\
GreenPhase (Ours) & 240K & 11.44 ($\sim$1.1$\times$) & 26.98 ($\sim$1.1$\times$) \\
GreenPhase (Ours) & 1.2M & 20.59 ($\sim$2.0$\times$) & 48.57 ($\sim$2.0$\times$) \\
\midrule
EQTransformer & 1.2M & 31.11 ($\sim$3.0$\times$) & 61.13 ($\sim$2.5$\times$) \\
\bottomrule
\end{tabular}
\end{table*}

\begin{table*}[t]
\centering
\caption{Comparison of carbon footprint and energy consumption in model training.}
\label{tab:carbon_train}
\begin{tabular}{lccc}
\toprule
\textbf{Model} & \textbf{Training Size} & \textbf{CO\textsubscript{2}e (kg)} & \textbf{Energy (kWh)} \\
\midrule
GreenPhase (Ours) & 60K & \textbf{0.189 (1.0$\times$)} & \textbf{0.445 (1.0$\times$)} \\
GreenPhase (Ours) & 240K & 0.747 (4.0$\times$) & 1.76 (4.0$\times$) \\
GreenPhase (Ours) & 1.2M & 1.80 (9.5$\times$) & 4.25 (9.6$\times$) \\
\midrule
EQTransformer & 1.2M & 72.04 (381$\times$) & 169.92 (382$\times$) \\
\bottomrule
\end{tabular}
\end{table*}

\section{Discussion}
The results demonstrate that the proposed Green Learning framework offers a compelling alternative to traditional deep learning, achieving performance comparable to the state-of-the-art model but with exceptional efficiency. This advantage stems from GreenPhase's core design: the GL framework is trained in a feed-forward manner, without relying on computationally expensive backpropagation. As a result, all parameters are learned in a single pass, leading to significantly faster training times and dramatically lower energy consumption. This inherent efficiency extends to inference. GreenPhase requires much smaller FLOPs. Furthermore, the model shows high data efficiency, delivering robust performance even when trained on a fraction of the data. These combined advantages in performance, training cost, and data robustness underscore the practical value and sustainability of the GreenPhase model.

\begin{figure}[h]
\centering
\includegraphics[width=\columnwidth]{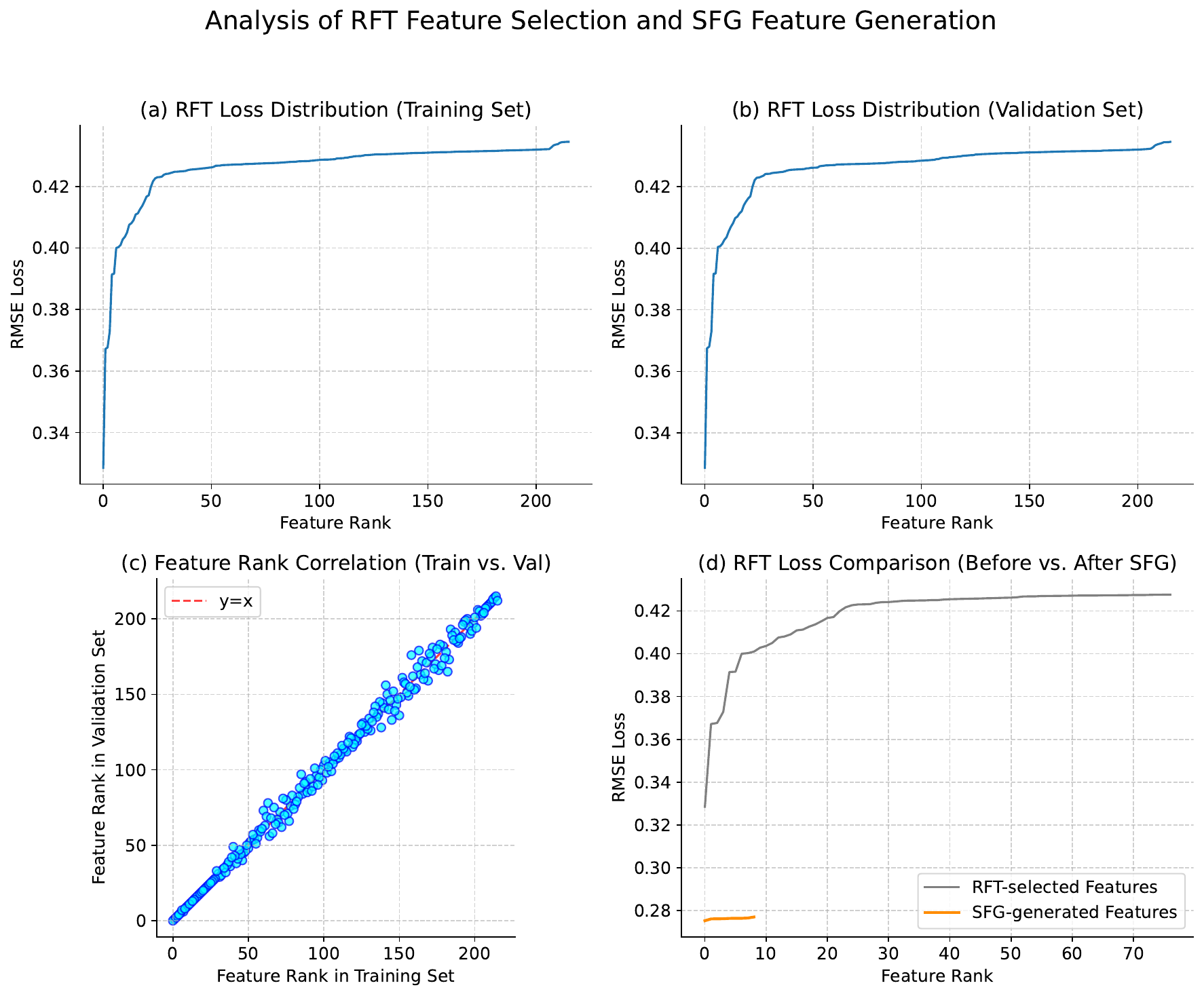}
\caption{Analysis of the RFT and SFG feature engineering steps for the Level 3 P-wave model. 
(a, b) The sorted RFT loss on the training and validation sets, respectively, shows that only a small fraction of features before the "elbow" are highly discriminative. 
(c) The high correlation of feature ranks between the training and validation sets confirms the stability of the RFT selection process. 
(d) The RFT loss of the new SFG-generated features is substantially lower than that of the original RFT-selected features, demonstrating the effectiveness of SFG. }
\label{fig:rft_sfg_analysis}
\end{figure}

The effectiveness of the key feature engineering components, RFT and SFG, is illustrated in Figure~\ref{fig:rft_sfg_analysis}, using the P-wave model at the coarsest resolution (Level 3) as an example. The first component, RFT, evaluates the discriminative power of each feature dimension generated by the Saab transform to select the most informative ones. Figures~\ref{fig:rft_sfg_analysis}(a) and (b) show the sorted RFT loss (RMSE) for all candidate features on the training and validation sets, respectively. The distinct "elbow" shape of the curves reveals that only a small fraction of features have low loss and are thus considered highly discriminative. Based on this observation, only features located before this elbow point are selected and used in subsequent processing stages.

To confirm the stability of this selection process, we compare the feature rankings between the two sets. Figure~\ref{fig:rft_sfg_analysis}(c) shows this rank correlation in a scatter plot. The strong diagonal distribution, which closely follows the $y=x$ reference line, demonstrates that the feature importance is highly consistent between training and validation, confirming the robustness of the RFT selection method.

The second key component, SFG, plays a crucial role in generating even more discriminative features from the RFT-selected candidates. Its effectiveness is shown in Figure~\ref{fig:rft_sfg_analysis}(d), which compares the sorted loss of the original RFT-selected features against the loss of the new SFG-generated features. The newly generated features (orange line) exhibit substantially lower loss values, indicating that SFG not only preserves but also enhances the discriminative information from the previous stage. This reduction in loss demonstrates the effectiveness of SFG in capturing higher-order dependencies among features, which are often overlooked when treating dimensions independently.

\begin{figure}[h]
    \centering
    \includegraphics[width=\columnwidth]{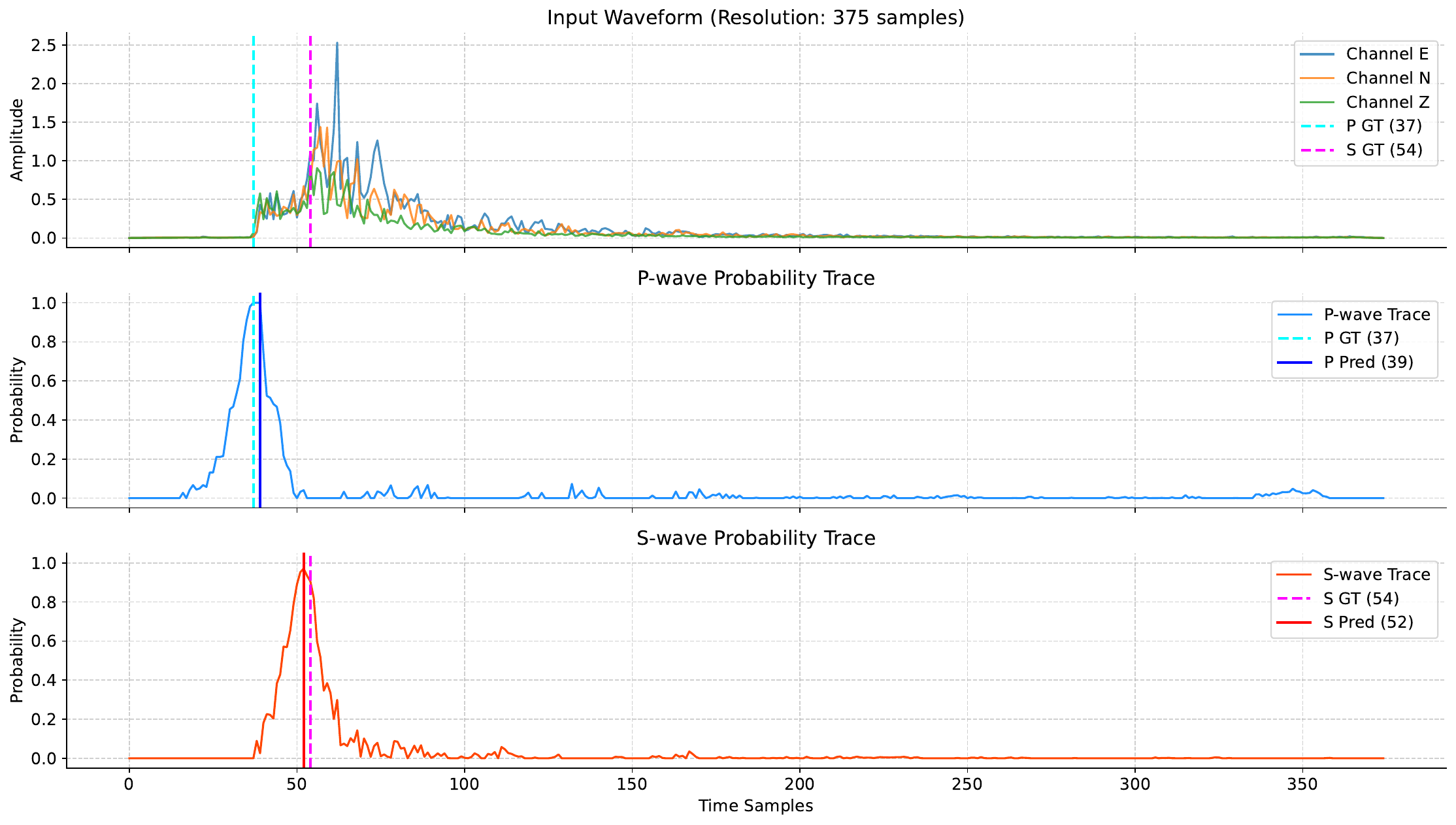}
    \caption{Prediction results at the coarsest resolution (Level 3, 375 samples).}
    \label{fig:level3_results}
\end{figure}

\begin{figure}[h]
    \centering
    \includegraphics[width=\columnwidth]{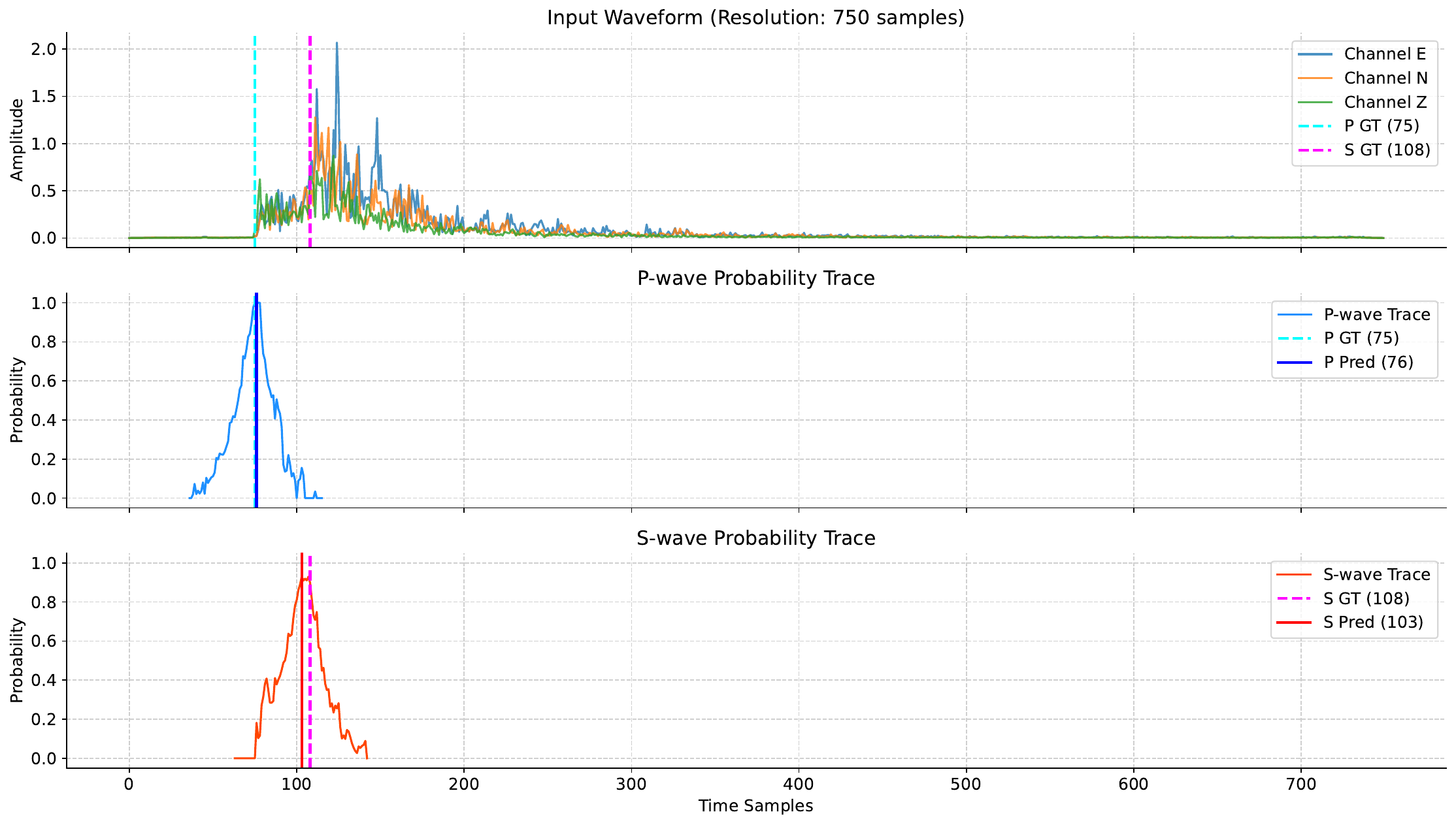}
    \caption{Refined prediction results at the intermediate resolution (Level 2, 750 samples).}
    \label{fig:level2_results}
\end{figure}

\begin{figure}[h]
    \centering
    \includegraphics[width=\columnwidth]{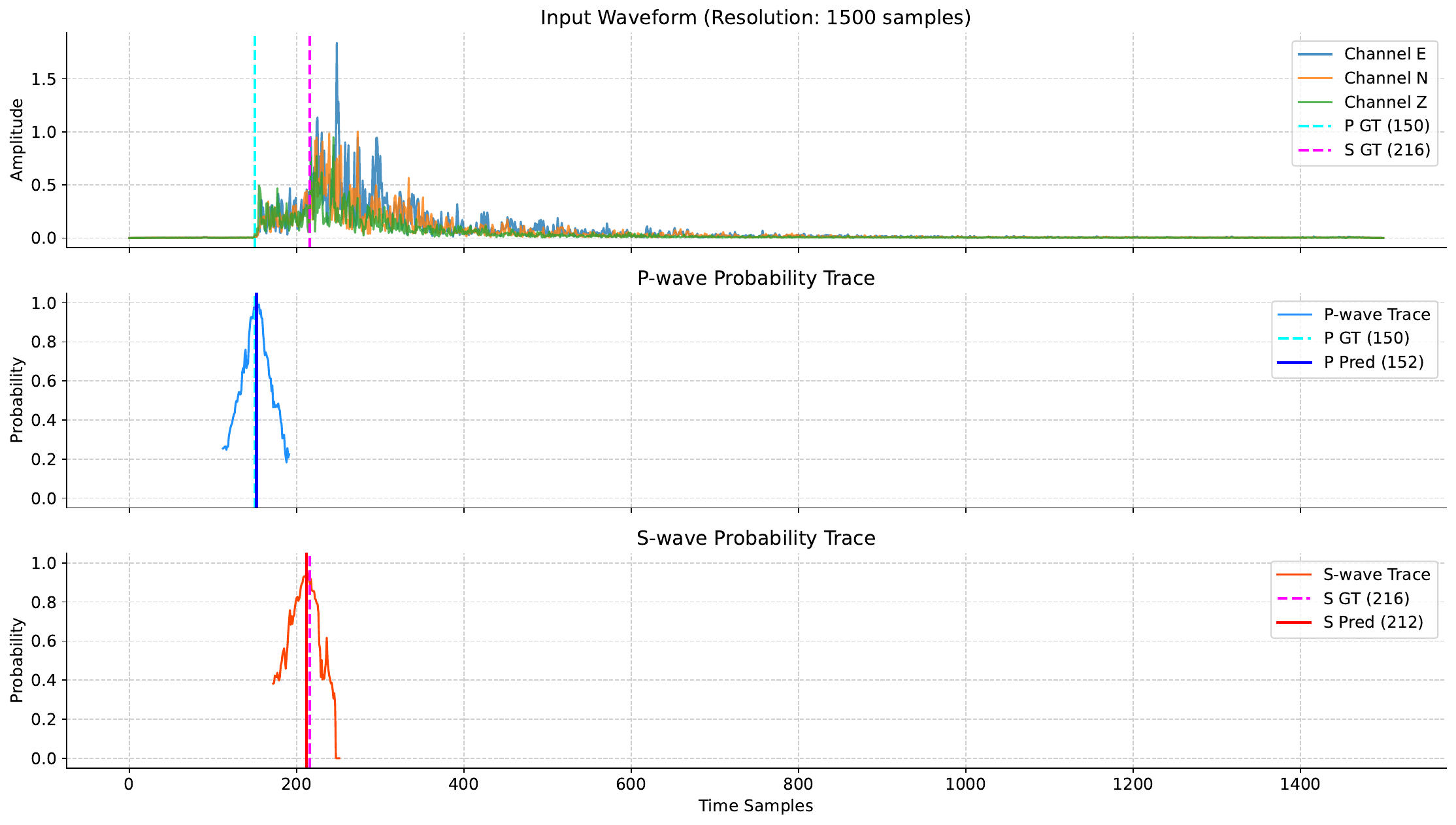}
    \caption{Final prediction results at the finest resolution (Level 1, 1500 samples).}
    \label{fig:level1_results}
\end{figure}

An important architectural innovation of GreenPhase is the multi-resolution, coarse-to-fine strategy, which directly addresses the inherent trade-off between high-resolution accuracy and computational efficiency in time-series analysis. Instead of applying computationally expensive feature extractors across a long, high-resolution sequence, GreenPhase first identifies a coarse Region of Interest (ROI) at a significantly downsampled resolution (Level 3). The effectiveness of this approach is visually evident in the results shown in Figures~\ref{fig:level3_results}, \ref{fig:level2_results}, and \ref{fig:level1_results}. This initial, coarse estimate allows the subsequent, finer levels to concentrate their analysis on a dramatically reduced search space. For instance, at the finest resolution (Level 1), the model only needs to analyze a narrow window of approximately 80 time locations. This represents a more than 18-fold reduction in the search space compared to naively scanning the full 1500-sample sequence. This hierarchical focusing of computation is a primary reason our model achieves high efficiency without sacrificing picking accuracy. The strategy is not only scalable for even longer time series but is also crucial for enabling real-time seismic monitoring on resource-constrained hardware, where processing an entire high-resolution waveform is often infeasible.

In terms of performance, it is worth noting that the F1 score for S-phase picking is slightly lower than that for P-phase picking. This performance gap is consistent with seismological observations, as S-wave picking is generally more challenging due to less impulsive arrivals and higher susceptibility to noise. Despite these challenges, our model demonstrates considerable robustness in detecting S-phases, achieving top recall and maintaining strong overall accuracy even with limited training data.

While GreenPhase demonstrates performance highly comparable to state-of-the-art models, we acknowledge a slight gap in the final F1 scores. This can be viewed as a trade-off for the substantial gains in computational efficiency, energy sustainability, and data robustness that our framework provides. Future work could aim to bridge this performance gap, potentially by exploring more advanced Green Learning modules or hybrid approaches, without significantly compromising the model's efficiency. Additionally, the current study was conducted exclusively on the pre-segmented STEAD dataset. A crucial next step is to validate the model's performance on continuous, real-time data from diverse regional seismic networks to assess its generalization capabilities in more challenging operational environments. Looking forward, the model's interpretable and modular nature provides a strong foundation for expanding its functionality to other critical seismological tasks, such as magnitude estimation and event classification, paving the way for a comprehensive and transparent analysis pipeline.

\section{Conclusion}

We have presented GreenPhase, a multi-resolution and feed-forward machine learning model for earthquake detection and phase picking built upon the Green Learning (GL) framework. By replacing computationally expensive backpropagation with a series of mathematically interpretable modules, our approach achieves performance comparable to state-of-the-art deep-learning models while dramatically reducing computational cost, model size, and training energy. The model's high efficiency makes it uniquely suitable for processing large-scale continuous seismic data, and its transparent design provides a fully traceable and trustworthy tool for geophysical analysis. Ultimately, this work demonstrates that the GL framework offers a powerful, efficient, and sustainable path forward for tackling large-scale challenges in the earth sciences.

\section*{Acknowledgments}
This research was financially supported by the National Science and Technology Council (NSTC). We are also grateful to the developers of the Stanford Earthquake Dataset (STEAD) for providing the open-access seismic waveform data essential to this study.

\section*{Computer Code Availability}

Name of the library: GreenPhase

Contact: yixingwu@usc.edu

Program language: Python
 
Software required: Python

Program size: 89KB

The source codes are available for downloading at the link: \url{https://github.com/star-wyx/GreenPhase}

\clearpage

\bibliographystyle{plainnat}
\bibliography{ref}

\end{document}